\documentclass[twocolumn]{bmcart}

\usepackage[utf8]{inputenc} 
\usepackage{color}
\usepackage{array}
\usepackage{booktabs}
\usepackage{amsthm}
\usepackage{amsmath,amssymb}
\usepackage{mathtools}
\usepackage{cancel}
\usepackage{framed}
\usepackage{graphicx}
\usepackage{subfigure}
\usepackage{appendix}
\usepackage{ulem}
\usepackage{verbatim}
\usepackage{hyperref}

\startlocaldefs

\endlocaldefs

\begin{document} 
\begin{frontmatter}
\begin{fmbox}
\dochead{Research}

\title{Nanosurveyor: a framework for real-time data processing}

\author[
   addressref={uppsala},            
   noteref={},                        
   email={benedikt@xray.bmc.uu.se}   
]{\inits{BJD}\fnm{Benedikt J.} \snm{Daurer}$^*$}
\author[
   addressref={lbnl},            
   email={hkrishnan@lbl.gov}
]{\inits{HK}\fnm{Hari} \snm{Krishnan$^*$}}
\author[
   addressref={lbnl},
   email={}
]{\inits{TP}\fnm{Talita} \snm{Perciano}}
\author[
   addressref={uppsala},
   email={filipe@xray.bmc.uu.se}
]{\inits{FM}\fnm{Filipe R.N.C.} \snm{Maia}}
\author[
   addressref={lbnl},
   email={}
]{\inits{DAS}\fnm{David A.} \snm{Shapiro}}
\author[
   addressref={lbnl,math},
   email={}
]{\inits{JAS}\fnm{James A.} \snm{Sethian}}
\author[
   corref={lbnl},   
   addressref={lbnl},            
   email={
benedikt@xray.bmc.uu.se, hkrishnan@lbl.gov, smarchesini@lbl.gov.}
]{\inits{SM}\fnm{Stefano} \snm{Marchesini}}

\address[id=uppsala]{ 
  \orgname{Laboratory of Molecular Biophysics, Department of Cell and Molecular Biology, Uppsala University},  
  \city{Uppsala}, 
  \cny{SE} 
}
\address[id=lbnl]{
  \orgname{Lawrence Berkeley National Laboratory},
  \city{Berkeley, CA},
  \cny{USA}
}
\address[id=math]{
  \orgname{Department of Mathematics, University of California, Berkeley},
  \city{Berkeley, CA},
  \cny{USA}
}


\end{fmbox}

\date{\today}

\begin{abstractbox}
\begin{abstract}
Scientists are drawn to synchrotrons and accelerator based light
sources because of their brightness, coherence and flux. The rate of
improvement in brightness and detector technology has outpaced Moore's
law growth seen for computers, networks, and storage, and is
enabling novel observations and discoveries with faster frame rates,
larger fields of view, higher resolution, and higher dimensionality.
 Here we present an integrated software/algorithmic framework designed
 to capitalize on high throughput experiments, and describe the
 streamlined processing pipeline of ptychography data analysis. The
 pipeline provides throughput, compression, and resolution as well as
 rapid feedback to the microscope operators.
\end{abstract}


\end{abstractbox}
\end{frontmatter}

\section{Introduction}

When new drugs are synthesized \cite{drugs}, dust particles are brought
back from space \cite{stardust}, or new superconductors are
discovered \cite{superconductor}, a variety of sophisticated X-ray
microscopes, spectrometers and scattering instruments are often
summoned to characterize their structure and properties. High
resolution and hyperspectral X-ray imaging, scattering and tomography 
instruments at modern
synchrotrons are among the workhorses of modern discovery to study
nano-materials and characterize chemical interactions or electronic
properties at their interfaces.


A new generation of microscopes are being pioneered, commissioned and
planned at several U.S Department of Energy (DOE) user
facilities \cite{ ptychonsls, winarski2012hard,
shapiro2013development} and elsewhere to achieve superior resolution
and contrast in three dimensions, encompassing a macroscopic field of
view and chemical or magnetic sensitivity, by coupling together the
brightest sources of tunable X-rays,  nanometer
positioning, nanofocusing lenses and faster detectors. 
Existing soft X-ray detector
technology in use at the Advanced Light Source (ALS) for example
generates 350 MBytes/second per instrument~\cite{fastccd}, commercial
detectors for hard X-rays can record 6 GB/second or raw data per detector
~\cite{pilatus,eiger}, and a synchrotron light source can support
40 or more experiments simultaneously 24 hours a day.  Furthermore,
accelerator technology such as multi-bend achromat \cite{max4} will increase
brightness by two orders of magnitude around the globe ~\cite{SRU,ultimate}.

Transforming massive amounts of data into the sharpest images
ever recorded will help mankind understand
ever more complex nano-materials, self-assembled devices, or to study
different length-scales involved in life - from macro-molecular
machines to bones - where observing the whole picture is as
important as recovering the local arrangement of the components. In
order to do so, there is a need for reducing raw data into meaningful
images as rapidly as possible, using the fewest possible computational
resources to sustain ever increasing data rates.


Modern synchrotron experiments often have quite complex processing pipelines, 
iterating through many different steps until
reaching the final output. 
One example for such an experiment is ptychography 
\cite{rodenburg,rodemburgprl07,Thibault2008}, 
which enables one to build up very large images by 
combining the large field of view of a high precision
scanning microscope system with the resolution provided by diffraction
measurements. 

Ptychography uses a small step size relative to the size of the
illuminating beam when scanning the sample, continuously generating
large redundant datasets that can be reduced into a high resolution
image. Resolution of a ptychography image does not depend on the size
or shape of the illumination. X-ray wavelengths can probe atomic and
subatomic scales, although resolution in scattering experiments is
limitated by other factors such as radiation damage, exposure and
brightness of the source to a few nanometers except in special  cases
(such as periodic crystals). 
 To reconstruct an image of the object from a series of
x-ray scattering experiments, one needs to solve a difficult phase
retriaval problem because at short wavelengths it is only possible 
to measure the intensity of the photons on a detector. 
The phase retrieval problem is made tractable in ptychography by
recording multiple diffraction patterns from overlapping regions of
the object, providing redundant datasets to compensate for the lack of
the phase information. The problem is made even more challenging in
the presence of noise, experimental uncertainties, optical
aberrations, and perturbations of the experimental geometry which
require specialized solvers and software \cite{Nashed2014,ptypy,
Marchesini2016}.

 In addition to its complex reconstruction
pipeline, a ptychography experiment involves additional I/O operations
such as calibrating the detector, filtering raw data, and
communicating parameters (such as X-ray wavelength, scan positions,
detector distance and flux or exposure times) to the analysis infrastructure.


Large community driven projects have developed frameworks optimized
for distributed data stream processing. Map-Reduce based solutions
such as Hadoop \cite{Hadoop, White:2009} and Spark \cite{Zaharia:2010}
provide distributed I/O, a unified environment, and hooks for running
map and reduce operations over a cloud-based network. Other frameworks
such as Flink \cite{Flink}, Samza \cite{Samza}, and Storm \cite{Storm}
are more tailored for realtime stream processing of tasks executing a
directed acyclic graph (DAG) \cite{DAG} of operations as fast as
possible. Workflow graphs such a Luigi \cite{Luigi} and Dask
Distributed \cite{Dask:2016, matthew:2015} provide an iterative
component, but are either optimized for batch processing and workers
are treated as a singular entity able to execute the DAG in its
entirety. 

Such frameworks target operations as a unit of tasks and generalize
the notion of resources, however the ecosystem is harder to
decentralize. These paradigms are not easily mappable to a production
beamline environment, where data from a detector might be running on a
field-programmable gate array (FPGA), the motion control system on a
real-time MCU, the acquisition control on a windows operating
system, and the scientist a macOS laptop. 
 The rest of the pipeline tasks might hop to several different
architectures including CPUs for latency bound tasks, and GPUs for high
throughput image processing and visualization.  While frameworks such as
Flink along with Kafka \cite{Kafka} (high throughput distributed
message system) and ZooKeeper \cite{ZooKeeper} (distributed
coordination and management) can be adopted to fit the described
processing environment, our solution at a lower level accomplishes the
same task with less computational and human resources.

 \textsc{Nanosurveyor} is a modular framework to support
distributed real-time analysis and visualization of data. The
framework makes use of a modular infrastructure similar to
Hummingbird \cite{hummingbird} developed to monitor flash x-ray imaging
experiments at free electron lasers (FELs) with high data rates 
in real time over multiple cores and nodes.

Within this framework, 
we developed a streamlined processing pipeline
for ptychography which unifies all components 
involved and allows users to monitor and quickly act
upon changes along the experimental and computational pipeline.

\section{Real-time Streaming Framework}

\textsc{Nanosurveyor} was developed to provide real-time feedback through analysis and visualization for
experiments performed at synchrotron facilities, and execute a complex
set of operations within a production environment. Its design is such
that it can be effectively adapted to different beamline
environments. It is built around a client-server infrastructure
allowing users to use facility resources while located at a beamline
or remotely, operating on live data streamed from the
beamline. Additionally, one can use the \textsc{Nanosurveyor} user
interface for off-line processing of experimental data saved on disk.
In this section we describe the resources and capabilities provided by
the modular streaming infrastructure.

\subsection{Modular Framework}

\begin{figure}
  \begin{center} \includegraphics[width=.95\columnwidth]{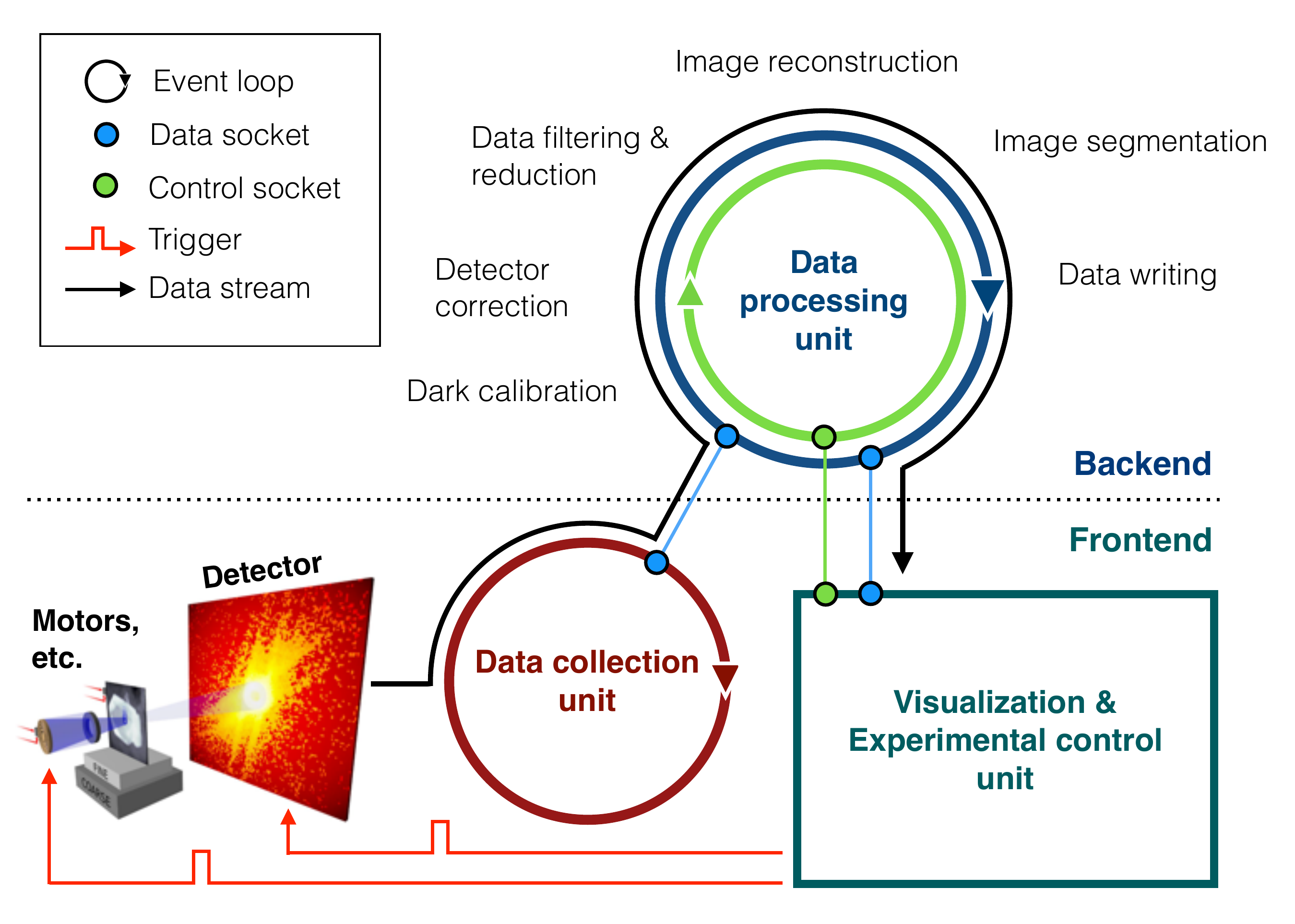}\end{center} 
  \caption{\label {fig:pipeline} Overview of the real-time
  streaming framework of Nanosurveyor. 
 The modular server-client infrastructure is divided into a back-end
  (running the data processing unit) and a front-end (running the visualization and control unit). 
  Once an experiment has started, the data collection unit 
  continuously receives new data packets from a detector and sends 
  raw data frames to the data processing unit. Depending on the 
  specific needs of the experiment, data is being corrected, 
  reduced, reconstructed and various outputs are written to 
  file. At all times, there is an active connection (asynchronous socket communication) 
  between all components (including the visualization interface)
  allowing the user to monitor progress while data is still being acquired and processed.}
\end{figure}

As described above, \textsc{Nanosurveyor} is designed to be adaptable and modular. 
Therefore, we designed it with a client-server infrastructure
(\ref{fig:pipeline}) enabling users to run their experiment while at
the beamline or remotely from their institution. This strategy also
allows the client to be very light and flexible  while the
server can be scaled according to the resources needed.

The \textsc{Nanosurveyor} infrastructure equips each module with two
fundamental capabilities. First, a description format language of
key-value pairs allows every module to describe its input and
output. Second, it provides the ability to describe the connection between 
the modules, including the front-end.

The capability to connect the communication path between modules
allows the end-to-end pipeline to be constructed and described
seamlessly. This is done through a proxy communication layer allowing
the modules to run either closely together or on completely separate
machines. This strategy is transparent to the beamline user and
accommodates both environments with centralized resources as well as
those where resources are spread across a network.

Additionally, as each module in the pipeline can be executed in its
own environment, \textsc{Nanosurveyor} provides dynamic parallelism by
allowing the user to scale the number of resources available to each
step: this is done by treating each stage as a worker process that can
be scaled up or down to address bottleneck or performance issues.

\subsection{Software Stack}

The core components of the \textsc{Nanosurveyor} streaming software
are written in Python using \texttt{ZeroMQ}, a high performance
messaging library \cite{zeromq} for network communication,
\texttt{PyQt4} \cite{pyqt2016} and
\texttt{PyQtGraph} \cite{Campagnola2016} for the graphical user
interface (GUI) and visualization, and \texttt{Numpy} \cite{Walt2011}
together with \texttt{Scipy} \cite{Jones2001} for manipulation of data
arrays. For some components, we used \texttt C extensions in order to
boost the performance to meet the demands of producing a real-time
interactive tool running at the beamline.

Python is a language with a robust and active community with libraries
that are well tested, supported, and maintained. Additionally, the
choice of Python allows our infrastructure to be flexible to the
demands of varying requirements of different processing pipelines. The
ptychography pipeline (discussed in detail later in the paper)
contains GPU optimized code and Python binding support easily allows
the \textsc{Nanosurveyor} infrastructure to provide support for these
types of hybrid architectures. The framework currently runs on Mac,
Linux and Linux-based cluster environments, and can be extended to
Windows platforms depending on support for module dependencies. The
core components that \textsc{Nanosurveyor} depends on are available on
all major platforms.

\subsection{Communication}

A critical component in generating usable real-time pipelines relies on 
the communication infrastructure. This enables a clear and
concise separation of the inputs and outputs at the module
level. Furthermore, it defines how modules communicate from beginning
to end, and ensures that tasks are load-balanced to achieve the appropriate
performance characteristics of the pipeline.

The communication in \textsc{Nanosurveyor} uses Javascript Object Notation (JSON) \cite{json}, 
an industry standard way of conveying metadata between modules as well as between the front-end and back-end. 
The metadata provides a human readable component.

\texttt{ZeroMQ} provides the communication backbone of the \textsc{Nanosurveyor} infrastructure. 
Using the publisher-subscriber model for the core components enables \textsc{Nanosurveyor} 
to provide a load-balancing scheme, which uses a backlog queue to avoid losing data when sufficient resources 
are not be available. The execution pipeline creates a command port and a data port. 
The command port allows metadata to reach and update parameters as well as return responses to keep 
status requests alive and provide feedback on the current state of the running module. 
The data port moves data through the pipeline, running the actionable item within each 
module and moving the result to the output queue to be processed by 
the next stage of the pipeline.

Two types of configurations are required: front-end and back-end. 
The front-end sets up the variables necessary for each module to function while 
the back-end configuration is responsible for allocating resources, 
balancing the load of workers, scheduling activities, 
and communicating between modules while providing feedback to the front-end. 

These two components provide the \textsc{Nanosurveyor} infrastructure
with the information it needs to establish the relevant connections,
receive and send parameters to ensure proper configuration, and
introspect the state of parameters and data to provide visual feedback
to the user when running through the processing pipeline.




\section{Client-Server architecture}

The \textsc{Nanosurveyor} framework consists of an assortment of core
components that ensure that the front-end provides easy to use and
adaptable interface while the back-end is efficient, resilient, and
responsive. The individual processing modules are all based on the
same structure: an event loop runs routing data from the control and
data sockets, waiting for tasks, asking the handler for configuration
parameters (JSON string), and processing data (receiving/sending
through the data socket).

\subsection{Back-end}

The main back-end handler is running a big \texttt{ZeroMQ} event loop. 
The main task of the handler is to register the modules that run on the 
back-end and ensure data and control paths are appropriately connected 
up and running. It also does the following:

\begin{itemize}
 \item Launches all the processing modules as separate processes (single-core or MPI) 
   and keeps track of the jobs started. This can be done with a batch processing system 
   such as SLURM (or any other queuing system) or by launching separate python processes;
 \item Creates the sockets for streaming pipeline, which is a list of control and data 
   sockets communicating between the handler and all the processing modules as well as 
   the data collector and the interface;
 \item Runs the event loop, takes commands, deals out data packets and handles everything 
   in the back-end including user interruption and other control and configuration commands.
\end{itemize}

\subsection{Data Tracking} 

Tracking and ensuring the correctness of data is an important part of
the execution pipeline.  The \textsc{Nanosurveyor} framework provides
a module called \texttt{nscxwrite} which allows customized writing of
files at different stages of the data acquisition pipeline (raw,
filtered, and reconstructed). This capability provides several
benefits, such as assurances to users that data moves correctly from
module to module and is not corrupted along the way, as well as an
ability to debug an algorithm that is executed within a complex
sequence of events.

Furthermore, the ability to save intermediate data can be enabled or
disabled (for performance reasons or to reduce storage) as well as
customized. The framework also comes with a standalone script called
\texttt{nsraw2cxi}, that translates raw detector
data to processed CXI files, and a script to stream simulated FCCD data 
through the pipeline for testing. The data format of the output files
follows the CXI file format \cite{Maia2012}.

\subsection{Logging}

\textsc{Nanosurveyor} also provides a way to debug a complex pipeline through
logging of both the output and error channels which includes
communication between modules as well as output and error that arise
from within modules.

The output of all modules are piped to STDOUT and STDERR within the
file system running each process
(\texttt{\$HOME/.nanosurveyor/streaming/log/}).

This is a useful tool that invokes \texttt{tail -f} on the piped
out/err files, making it possible to monitor what is going within the
individual processing modules.

\subsection{Graphical User Interface}

For the front-end, the framework provides a versatile GUI based on
\texttt{PyQt4} and \texttt{PyQtgraph} for monitoring, visualizing and
controlling the data processed live or post-processed through the
pipeline. \texttt{PyQt4} (built on \texttt{Qt}) provides the ability
to construct and modify the user interface to easily add and remove
functionality while \texttt{PyQtgraph} provides access to advanced
visualization functionality for data that can be represented as images
or volumes. Several common operations provided through the framework
include:
 
\begin{itemize}
 \item View the content of already processed files: Inspect reconstructions from collected data and provide 
   other useful utilities  (histograms, error line plots, correlation plots, and others); 

 \item Control and monitor the streaming: Configure streaming, inspect live reconstruction, 
  monitor performance (upload/download rates, status update of the streaming components);

 \item Simulate an experiment starting from an SEM image or similar;

 \item Process and inspect, through a provided interface, data from custom modules processed on 
  the back-end (e.g. data from a ptychography or Tomography reconstruction).
\end{itemize}
 
Generally speaking, the design facilitates adding new modules to the GUI, e.g. 
a viewer for tomograms or similar. This flexibility allows the front-end to be customized 
for different beamline processing environments. 


Finally, the architecture aims to be modular in the front and back-end of the 
client-server architecture, meaning that there is a template structure for the 
basic features of a processing module. Additionally, in principle, any given 
processing module can be hooked into this network (e.g. tomography, spectral 
analysis, or any other image analysis). 

\section{Streaming ptychography}

\begin{figure}
  \begin{center} \includegraphics[width=.95\columnwidth]{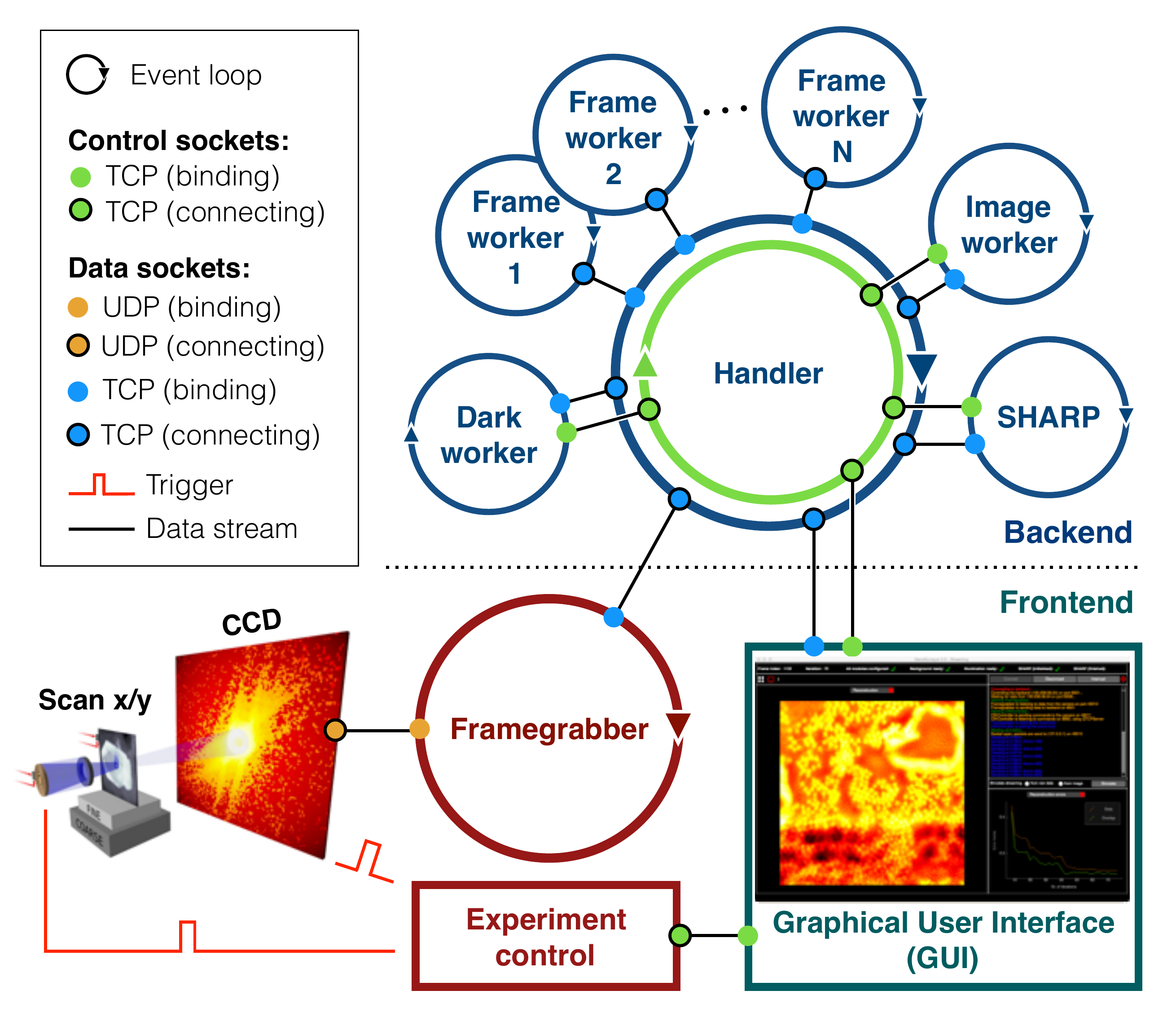}\end{center} 
  \caption{\label {fig:pipeline_als} Streaming pipeline implemented 
    at the ALS for ptychographic imaging. The software structure follows the 
    same logic as sketched in Figure \ref{fig:pipeline}. 
    Once a new scan has been triggered by the
    experimental control, a frame-grabber continuously receives raw data
    packets from the camera, assembles them to a frame and sends raw
    frames to the back-end. Incoming frames are processed by different
    (and independent) workers of the back-end and reduced data
    is send back to the
    front-end and visualized in a graphical user interface (GUI). 
    A handler is coordinating 
    the data and communication workflow. 
 }
\end{figure}

We adapted the outlined streaming framework described above for the specific 
needs of ptychography and are currently implementing this ptychography streaming 
pipeline at the beamline for scanning transmission X-Ray microscopy (STXM) at
the Advanced Light Source (ALS). The main motivation for this project is to make 
high-resolution ptychographic reconstructions available to the user in real-time. 
To achieve this goal, we streamlined all relevant processing components
of ptychography into a single unit. A detailed outline of our pipeline is sketched 
in Figure \ref{fig:pipeline_als}.

As described in the previous sections, we follow the idea of a modular
streaming network using a client-server architecture, with a back-end
for ptychographic processing pipeline and a front-end for
configuration, control and visualization purposes.

On the back-end side, the streaming infrastructure is composed of a
communication handler and four different kinds of workers addressing
dark frames, diffraction frames, reduced and downsampled images and
the ptychographic reconstruction using a software package for scalable 
heterogeneous adaptive real-time ptychography (\textsc{SHARP}
 \cite{Marchesini2016}).  
The handler bridges the back-end
with the front-end and controls the communication and data flow among
the different back-end workers. The dark worker accumulates dark
frames and provides statistical maps (mean and variance) of the noise
structure on the detector. The frame workers transform raw into clean
(pre-processed) diffraction frames. This involves a subtraction of the
average dark, filtering, photon counting and downsampling. Depending
on the computing capacities of the back-end, it is possible to run as many
frame workers simultaneously as needed. 
The image worker reduces a collection of clean diffraction frames, 
producing low-resolution image
reconstructions and an initial estimate of the illumination function
which, together with the clean diffraction frames, is then feeded as an
input for the high-resolution ptychograhic reconstruction worker
(\textsc{SHARP}).

\begin{figure}
 \begin{center}\includegraphics[width=.95\columnwidth]{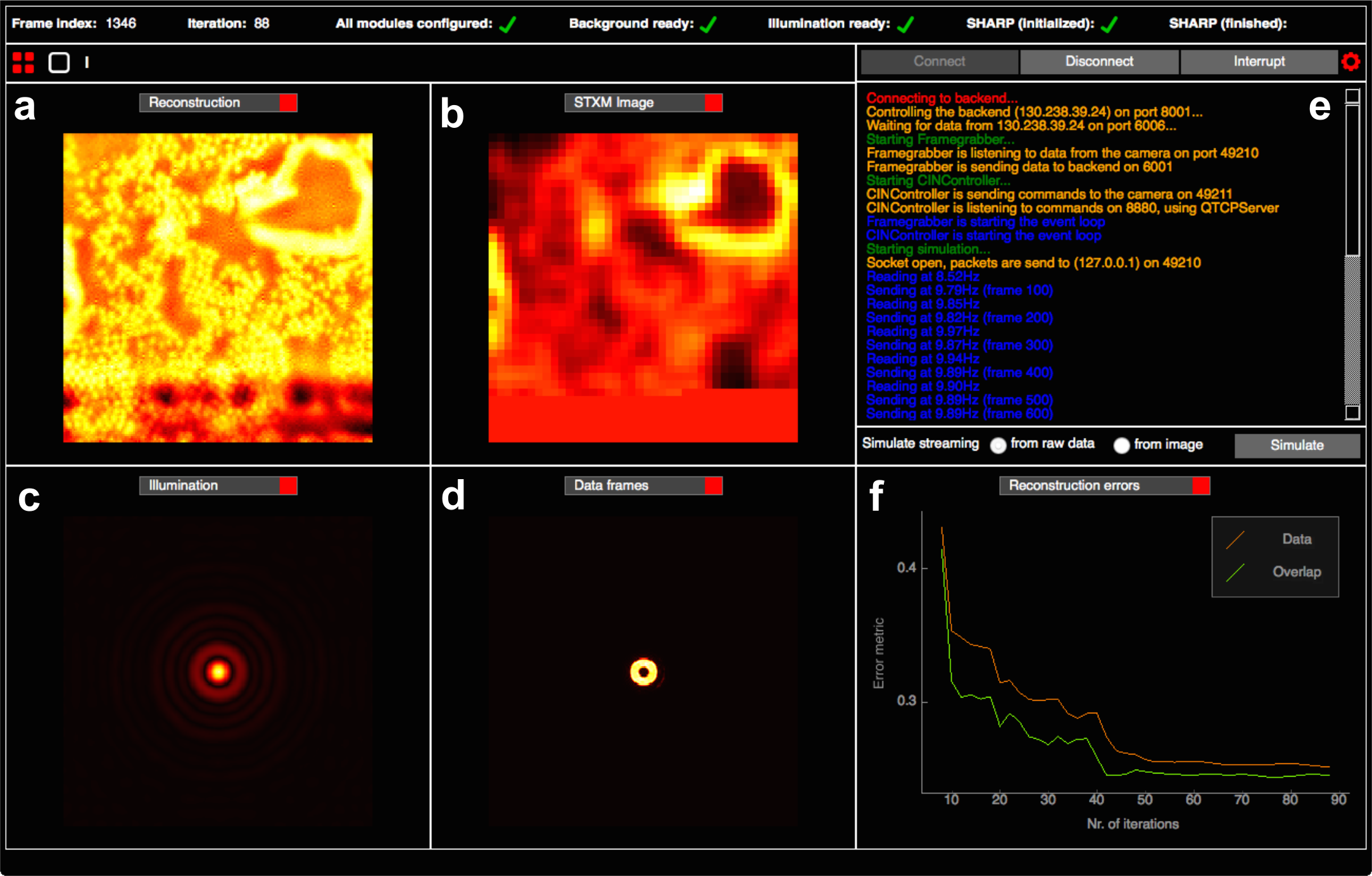}\end{center}
 \caption{\label {fig:interface}
   Graphical User Interface (GUI) for the ptychographic streaming pipeline 
   implemented at the ALS. The interface provides
   (a) Real-time view of the ptychographic reconstruction (high resolution),
   (b) Real-time view of the STXM analysis (low resolution),
   (c) Current guess of the illumination function,
   (d) Current processed data frame,
   (e) Logging and error messages and
   (f) Error metrics of the iterative reconstruction process, 
   and other control and monitoring elements around.
}
\end{figure}

The front-end consists of a worker that reads raw data frames from a fast charge-coupled device (FCCD) \cite{fccd}, 
coordinating with a separately developed interface for controlling the experiment 
(such as motors and shutters) and a graphical user interface (GUI) which is used both
for visualizing and controlling the ongoing reconstruction.
An example view of the GUI for streaming ptychography 
is shown in Figure \ref{fig:interface}.

Following the data flow along the streaming pipeline, the starting
trigger comes from the control interface which initiates a new
ptychographic scan providing information about the scan (step size,
scan pattern, number of scan points) and other relevant information
(e.g. wavelength) to the back-end handler. Simultaneously, the control
sends triggers to the scanning motors and the FCCD.  A typical
ptychographic scan combines the accumulation of a given number of dark
frames together with scanning the sample in a region of interest. The
frame-grabber, already waiting for raw data packets to arrive,
assembles the data and sends it frame-by-frame to the back-end
handler. When dealing with an acquisition control system that runs 
independently, the handler can distinguish between dark and
data frames using counters. Dark and data frames
 are distributed  to the corresponding workers. Having
clean diffraction frames and an initial guess for the illumination ready, the
\textsc{SHARP} worker is able to start the iterative reconstruction
process.
\textsc{SHARP} initializes and allocates space to hold all frames in a scan, 
computes a decomposition scheme, initializes the image and starts the
reconstruction process.  Unmeasured frames are either set to a bright-field frame (measured by removing the sample) or their weight is set to 0 
until real data is received.

Depending on the configuration, data at different states within the
streaming flow can be displayed in the GUI and/or saved to a CXI file
via the \texttt{nscxiwrite} worker module.

All components of the streaming interface run independent event loops and
use asynchronous (non-blocking) socket communication. To maximize performance, 
the front-end operates very close to the actual experiment,
while the back-end runs remotely on a powerful GPU/CPU cluster.

\subsection{Pre-processing of FCCD data}

We developed the following processing scheme for denoising and
cleaning the raw data from the FCCD and preparing frames for the
ptychographic reconstruction,

\begin{framed} 
\begin{enumerate} 
\item Define center (acquire some diffraction frames and compute the center of mass if needed). 
This is needed for cropping, and to deal with beamstop transmission;
\item Average dark frames: we first acquire a sequence of frames when
 no light is present, and compute the average and standard deviation
 of each pixel and readout block. We set a binary threshold to define
 bad (noisy) pixels or bad (noisy) ADC channels, when the standard
 deviation is above a threshold or if the standard deviation is equal
 to 0;
\item Remove offset using the overscan linear or quadratic offset:
stretch out the read-out sequence in time and fit a second order
polynomial over the overscan;
\item Identify the background by thresholding; 
\item Perform a Fourier transform of the readout sequence of 
the background for each channel, remove high frequency spikes by thresholding, and subtract from data;
\item Threshold signal below 1 photon;
\item Divide by beamstop transmission;
\item Crop image around center;
\item Downsample: take a fast Fourier transform (FFT), 
 crop or multiply by a kernel (e.g. Gaussian) and take inverse fast
 Fourier transform (IFFT).
\end{enumerate} 
\end{framed}

\subsection{Simulation}

For testing the functionality and performance of the streaming
ptychography pipeline as well as exploring different configurations,
we developed a protocol that simulates an entire ptychography scan.
Using a simulated illumination from a Fresnel Zone Plate (FZP) and
basic scan parameters (number of scan points, the scanning step size,
the scanning pattern), diffraction patterns from a well-known test
sample are calculated in the same raw data format as those generated
by the FCCD.  As a last step, Poisson noise and
a real background are added to the data. These raw data packets
together with the simulated metadata are introduced to the end-to-end
streaming pipeline and produce outputs as shown in
Figure \ref{fig:interface}.

One major benefit of this feature is the ability to scale and test the
pipeline at different acquisition rates and therefore be able to provide performance
metrics on the behavior of a sequence of algorithms enabling
developers to further improve their execution pipeline. 

In a simple performance test, we simulated a 40x40 scan producing 1\,600 raw data frames 
which were sent by a virtual FCCD at a rate of 10 Hz. At the end of the pipeline, we observed a complete reconstructed image 
after around 5 minutes. This translates into a streamlining pipeline rate of
about 2 Hz, with most of the time spent on filtering and
cleaning the individual frames. 
A significant portion of the pre-processing time is unique to the FCCD pipeline. 
While this rate is still far from ideal, it can be easily be sped up and scaled by using parallel
execution, load-balancing strategies, and eventually through high throughput GPU optimizations. 
With further improvements on the
performance on the individual components as well as optimization of
the network communication, we expect a substantial increase of the 
processing rate. 

\subsection{Experimental Data}

Experimental data produced by the FCCD can involve missing frames,
corrupted frames, and timing issues between different hardware and
software components. In addition, the correct choice parameter values
for the ptychographic reconstruction might be inherent to the data
itself and can thus carry from exeperiment to experiment.  To make
\textsc{Nanosurveyor} more robust for such cases, it is desirable to
expose configuration parameters as a runtime or heuristic feature
rather than determined them at execution time, and take a more
data-based approach where options are set based on feature detection.


\section{Considerations}
Performance considerations and additional limitations must be
understood and considered in integrating such an execution pipeline in
a production environment. While the following list is not
comprehensive, in building this environment, we have considered the
following:

\begin{itemize}
     \item {Limits (performance, algorithm, memory, disk) to software
     and hardware need to be considered. The \textsc{Nanosurveyor}
     infrastructure provides logging support while the \texttt{ZeroMQ}
     publisher-subscriber model allows a stuck or crashed process to
     be replaced with another. The current solution
     \textsc{Nanosurveyor} can be made more robust and this is work
     that is considered as active and ongoing;\label{limits}}
     
     \item {Hardware failures are inevitable in a production
     environment involving machinery. Recovery from these types of
     issues requires customization for each beamline
     environment. Within Nanosurveyor, there is a heartbeat for each
     module and a base mechanism within the framework to inform the
     user that a failure (or multiple failures) might have occurred;}
     
     \item {Interrupting experiments should be a core use case of any
     real-time feedback loop when trying to get an understanding of
     the data as quickly as possible. Once information about the
     material is flowing through the computational pipeline, it is
     valuable to be able to determine if an experiment is, in fact,
     failing or uninteresting. This can occur in many ways such as
     wrong setup, wrong material or wrong region of scanning. For
     these scenarios it is prudent for a working pipeline to be able
     to abort, clear out the pipeline, and reset itself;}
     
     \item {Expensive operations and algorithms executed in a beamline
     operating environment may have varying degrees of performance
     characteristics (see bullet point on Limits 
). These
     characteristics can often slow down the overall pipeline if any
     one of the operations is inefficient. \textsc{Nanosurveyor}
     attempts to get around this issue in two ways: first, it allows
     for a load-balancing approach where more workers can be added to
     the expensive stages of the pipeline. Second, using the
     \texttt{ZeroMQ} queue, the beamline can still operate with the
     slowdown and backlog while ensuring that the pipeline can
     continue to function, at least until hardware memory runs out.
     This issue can also be mitigated by evaluating the performance of
     the module and if possible optimizing the algorithm as well.}
\end{itemize}

\section{Conclusions \& Future Work}

This work introduced \textsc{Nanosurveyor} - a framework for real-time
processing at synchrotron facilities. The infrastructure provides a
modular framework, support for load-balancing operations, the ability to
run in a distributed client-server mode, and gives feedback on each
stage of a complex pipeline.

The framework was adapted to support streamlined pipelines for
ptychography. In this case, expensive stages such as pre-processing
are load-balanced with multiple workers, and image reconstruction are
parallelized over MPI to compute efficiently in a distributed manner.
Results from every stage of the pipeline are then transmitted to the
front-end, providing users at the beamline comprehensive knowledge of
the experiment and of how the data is transformed from start of
acquisition to end output. Although the \textsc{Nanosurveyor}
framework provides several core capabilities that are necessary for
operating at typical beamlines, there are several key advances that we
are currently working on to make the computational pipeline
complete. A couple of highlights include:

\paragraph{
Iterative execution, instrument control:} Adding support for
controlling the beamline itself will complete the current pipeline and
provide an iterative execution loop enabling future pipelines to
adaptively acquire and analyze data from the operating beamline, and
automatically request more data when necessary. For example, if the
reconstruction detects bad frames, or that the sample has drifted,
then more frames can be automatically requested on the fly without
interrupting the overall experiment. If the reconstruction determines
that part of the image being acquired is empty or uninteresting it
could request fewer frames and focus on the relevant part of the
sample.

\paragraph{
Optimizing pipeline execution:} Currently communication occurs over
\texttt{ZeroMQ} providing many benefits, including dealing with backlog, automated
load-balancing, and the ability to interleave work running different
stages of the execution pipeline. We are also investigating ways to
fuse modules to optimize execution times. Making communication
agnostic by using handles enables efficient use of memory optimization
strategies, socket communication, or saving on data movement costs
e.g., transferring data between GPU-based modules by moving a pointer
rather than copying data.


In conclusion, we have presented a framework that is built to run at
modern beamlines, can handle the geographic considerations between
users and experiments running at synchrotron facilities, and supports
real-time feedback. These features, along with the modular design,
provide a foundation that can be extended and readily deployed on many
of the beamlines in use today. Further information about
\textsc{Nanosurveyor} is available at \url{http://www.camera.lbl.gov/software}
or upon request to camera-nanosurveyor@lists.lbl.gov.

\begin{backmatter}
\section*{Competing interests}
  The authors declare that they have no competing interests.

\section*{Author's contributions}
BJD, HK, TP, FM and SM  designed and implemented the real-time streaming
framework. BJD, HK, TP, JAS and SM  wrote the manuscript with
contributions from all. DAS translated the preprocessing code from
matlab to python and helped us testing the streaming
ptychography framework at the ALS.

\section*{Acknowledgements}
This work was partially funded by the Center for Applied Mathematics
for Energy Research Applications, a joint ASCR-BES funded project
within the Office of Science, US Department of Energy, under contract
number DOE-DE-AC03-76SF00098, by the Swedish Research Council and by
the Swedish Foundation for Strategic Research.  The Advanced Light
Source is supported by the Director, Office of Science, Office of
Basic Energy Sciences, of the U.S. Department of Energy under Contract
No. DE-AC02-05CH11231.


\bibliographystyle{bmc-mathphys} 
\bibliography{references}


\end{backmatter}
\end{document}